\documentclass[sigconf]{acmart}
\usepackage{booktabs}
\usepackage{amsmath}
\usepackage{amsfonts}
\usepackage{enumitem}
\usepackage{amssymb}
\usepackage{balance}
\usepackage{subfig}
\usepackage{multirow, makecell}
\usepackage{gensymb}

\usepackage[utf8]{inputenc}
\usepackage[T2A,T1]{fontenc}
\usepackage[french,russian,polish,english]{babel}
\usepackage{CJKutf8} 
\DeclareMathOperator*{\argmax}{arg\,max}

\newfont{\mycrnotice}{ptmr8t at 7pt}
\newfont{\myconfname}{ptmri8t at 7pt}

\begin{document}
\copyrightyear{2018}
\acmYear{2018}
\setcopyright{acmlicensed}
\acmConference[DH'18]{2018 International Digital Health
Conference}{April 23--26, 2018}{Lyon, France}
\acmBooktitle{DH'18: 2018 International Digital Health Conference,
April 23--26, 2018, Lyon, France}
\acmPrice{15.00}
\acmDOI{10.1145/3194658.3194678}
\acmISBN{978-1-4503-6493-5/18/04}

\title{Hearts and Politics: Metrics for Tracking Biorhythm Changes during Brexit and Trump}

\author{Luca Maria Aiello}
\affiliation{%
  \institution{Nokia Bell Labs}
}
\email{luca.aiello@nokia-bell-labs.com}

\author{Daniele Quercia}
\affiliation{%
  \institution{Nokia Bell Labs}
}
\email{quercia@cantab.net}

\author{Eva Roitmann}
\affiliation{%
  \institution{Nokia}
}
\email{eva.roitmann@nokia.com}

\renewcommand{\shortauthors}{Aiello, Quercia, Roitmann}

\date{31 October 2017}

\begin{abstract}
Our internal experience of time reflects what is going in the world around us. Our body's natural rhythms get disrupted for a variety of external factors, including exposure to collective events. We collect readings of steps, sleep, and heart rates from 11K users of health tracking devices in London and San Francisco. We introduce measures to quantify changes in not only volume of these three bio-signals (as previous research has done) but also synchronicity and periodicity, and we empirically assess how strong those variations are, compared to random expectation, during four major events: Christmas, New Year's Eve, Brexit, and the US presidential election of 2016 (Donald Trump's election). While Christmas and New Year's eve are associated with short-term effects, Brexit and Trump's election are associated with longer-term disruptions. Our results promise to inform the design of new ways of monitoring population health at scale. 
\end{abstract}

\maketitle


\section{Introduction} \label{sec:introduction} 

Our body pulses in cycles: we sleep or waken, are hungry or full, are alert or tired. The most dominant period in a person's rhythms is the circadian cycle. Major departures from the normal range of the period have been associated with endogenous factors (e.g., illness) or exogenous ones (e.g., an external event inducing fear). Previous work has explored the relationship between circadian cycles and external factors, linking prolonged disruption of rhythms to pathological conditions, including cancer~\cite{saurbh11,takahashi08}. Nowadays, ``the alternation of sleep and walking and all the bodily cycles attendant on those states''~\cite{lynch1972time} can be measured based on the use of social media~\cite{golder11,Wang16,DeChoudhury17computational}, of augmented-reality games~\cite{althoff2016influence,graells2017effect},  and, more reliably, of activity trackers~\cite{althoff2017large,shameli2017gamification}. 

Yet, previous research has rarely ventured into: \textit{i)} studying activity metrics beyond their volume; and \textit{ii)} linking these metrics' changes to global events. This study aims at exploring these two aspects for the first time, and it does so by relying on large-scale data collected data from Nokia Health monitoring devices used by 11,600 customers who live in London (67\% of users) and San Francisco (33\%) over the course of 1 year (from 1st April 2016 to 30th April 2017). Our users are 44\% female, and their median age is 42 years. All users opted-in for research studies,  and their data has been processed in an anonymized form. We consider three types of \textit{activities}: total number of \textit{steps} walked during the day; \textit{sleep duration} measured in number of minutes slept at night; the estimated \emph{sleep time} when the user went to bed for the night (hour and minutes, adjusted for timezone); and the average \textit{heart rate} (beats per minute) measured once per day. Steps and sleep are measured by Nokia Health devices (e.g., former Withings wristbands and smart watches). All activities are measured at daily level for each user. For heart rates, when multiple measurements are available on the same day, we average them out. Our users represent a sample of the larger user population and are selected based on the number of days they used their devices: indeed, to reduce sparsity, we consider users who, for at least $90\%$ of the days, measured their heart rates. This leaves us with $3.8M+$ daily activity summaries.

By drawing from previous physiological and psychological studies, we derive metrics that relate walking, sleeping, and heartbeat to well-being. We characterize those three activities in terms of their \textit{volume} (the raw value of the signal over time, \S\ref{sec:volume}), \textit{synchronicity} (the degree to which the cycles of different people are in phase, \S\ref{sec:synchronicity}) and \textit{rhythm} (the activity periodicity, \S\ref{sec:rhythms}). We show how these metrics vary over the entire year, and how such variations represent distinctive signatures for four collective events: Christmas, New Year's Eve, Brexit, and the US election of 2016. We find that users slept more than usual during the Christmas period and, as one expects, slept less than usual during Brexit, the US election, and New Year's eve. Brexit and the US election are also associated with long-term disruptions in two main ways. First, in terms of sleeping patterns: users became heavily out-of-sync in the weeks after Brexit and even more so after the US election. Second, in terms of heart rates, we found major shifts in rhythm and volume, especially in the months around the US presidential election. These results suggest that our three metrics effectively capture how our biorhythms change during large-scale events, opening up new ways of monitoring population health at scale\footnote{Additional material is on \url{http://goodcitylife.org/health}}. 


\section{Volume} \label{sec:volume}

\begin{figure*}[t]
\centering
\includegraphics[width=.30\textwidth]{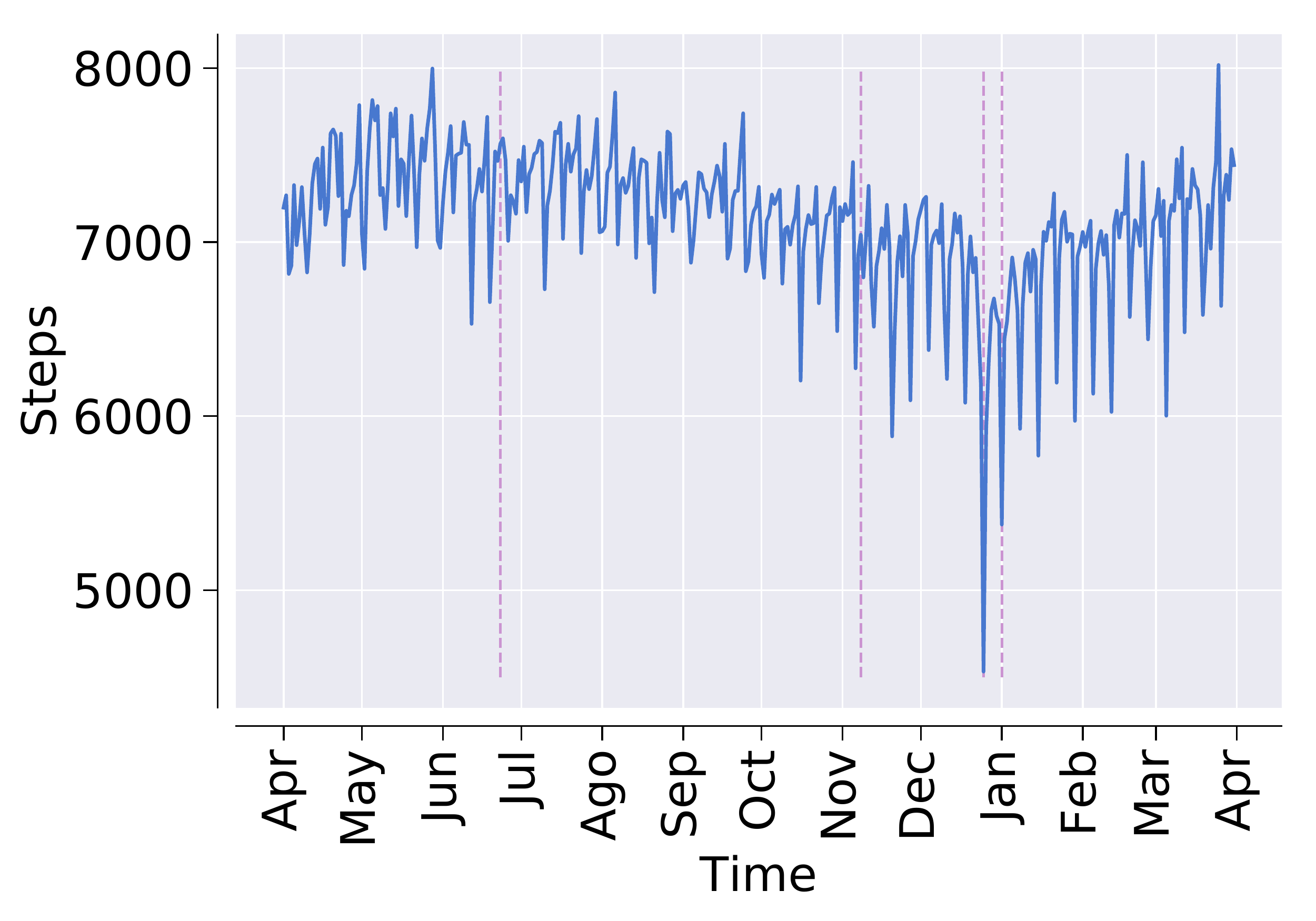}
\includegraphics[width=.30\textwidth]{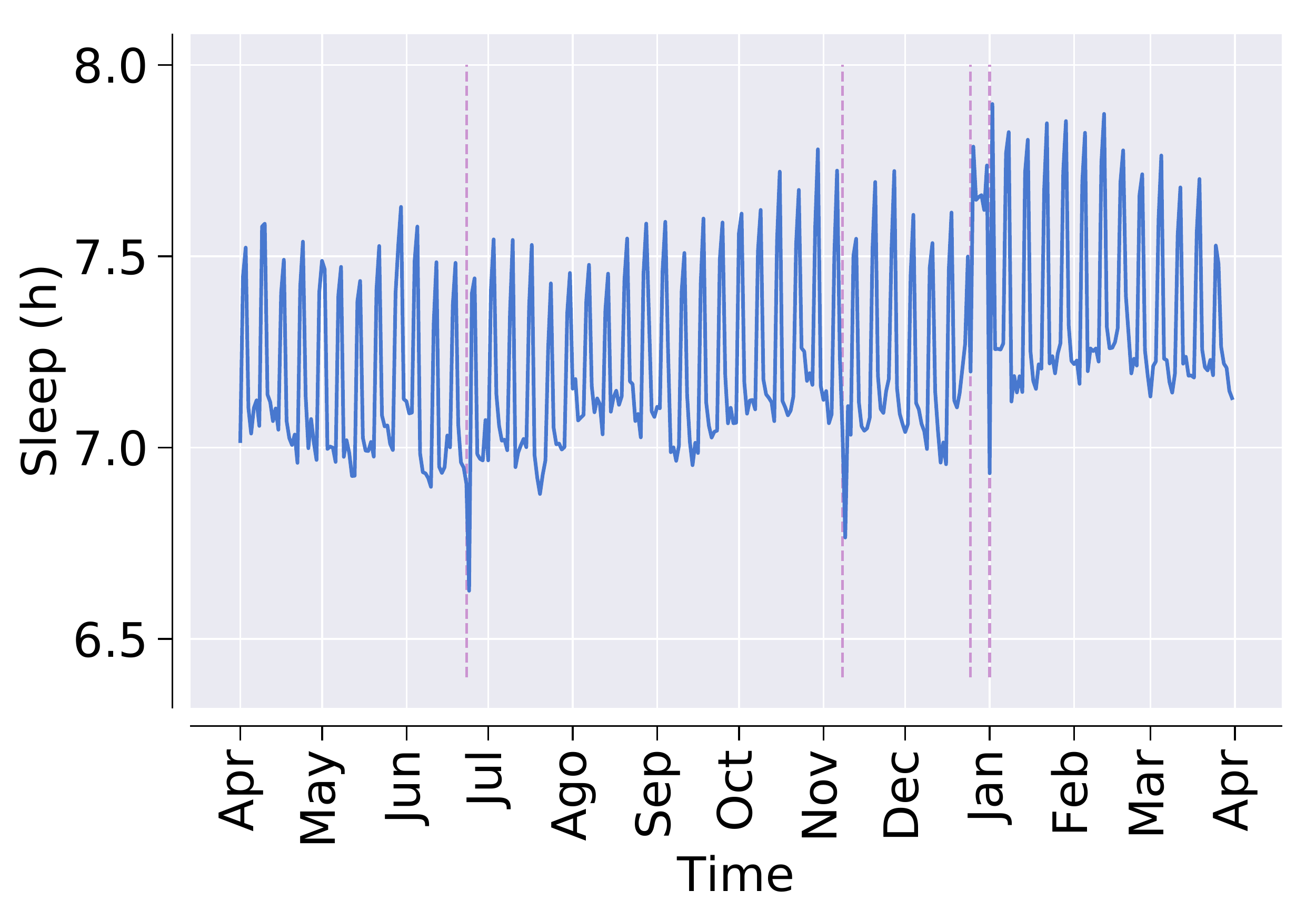}
\includegraphics[width=.30\textwidth]{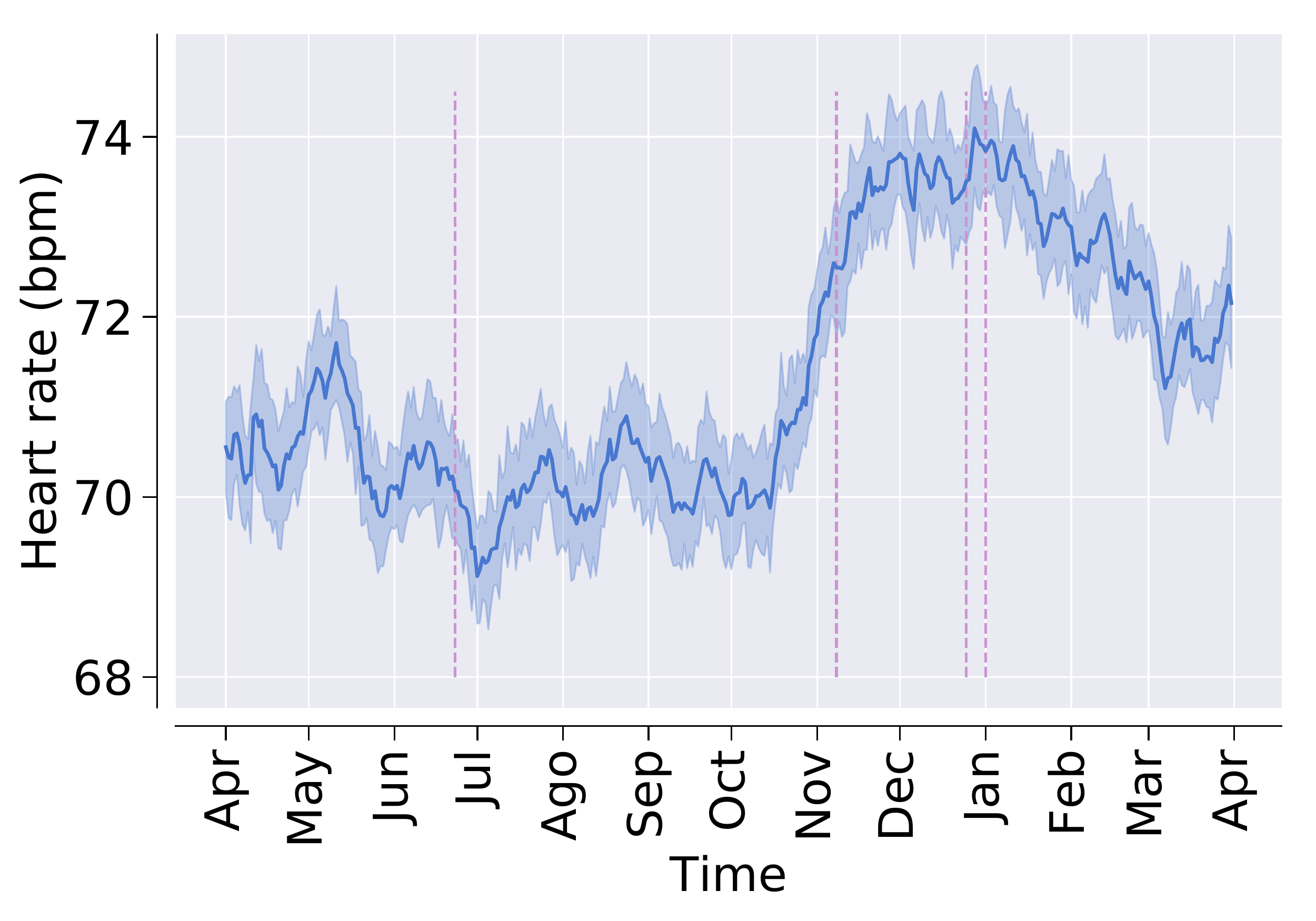}
\caption{Daily average number of steps, hours slept, and heart rate (with 95\% confidence interval). The dates of the following four events are marked by horizontal lines (left to right): the Brexit referendum, the US presidential election, Christmas and New Year's Eve.}
\label{fig:volume}
\end{figure*}

The amount of steps, hours of sleep and the dynamics behind heart rates have all been related to health outcomes. Physical activity boosts the levels of immune cells, and that results in a considerable reduction of sick days -- from children to elderlies~\cite{nieman00}. Sleep deprivation has been found to make people accident-prone on the road, unproductive at work~\cite{wiseman2014night}, and subject to brain aging~\cite{ferrie11,wiseman2014night}. Sleep deprivation also increases the chances of ailments such as hypertension~\cite{vetter16}, cancer~\cite{scott01}, diabetes, and obesity~\cite{reilly05}, and, as such, increases mortality rates~\cite{kripke02}. In this work, to capture the amount of steps, sleep, and heart rates, given the measurement of an activity $A$ (e.g., steps), we denote the activity of user $u$ on day $t$ with $A_u(t)$, and compute the average activity during day $t$ at population level as $\overline{A}(t) = \frac{\sum_{u \in U} A_u(t)}{|U|}$.

In Figure~\ref{fig:volume}, we plot the average daily number of steps, hours of sleep, and heart rates for the whole year. The plots of steps and hours of sleep are spiky, and that comes from our typical  weekly patterns: during weekends, people usually walk less and sleep more. Some of the spikes are much more prominent than the others though, and correspond to four major events (marked with dashed lines in the plots): the ``Brexit'' referendum in which the UK electorate voted to leave the EU on the 23th of June 2016, the US presidential election on the 8th of November 2016, and Christmas and New Year's Eve of the same year. For steps (first plot in Figure~\ref{fig:volume}), there are two low points, which  correspond to Christmas and New Year's Eve. For hours of sleep (second plot), there are three low points, which correspond to Brexit, the US election, and New Year's Eve.  Finally, for heart rates (third plot), there are a few peaks and low points but they are limited -- the most remarkable trend is represented by  a considerable collective increase of heart rate just around the US election.  These results might suggest that increases in heart rates are in a cause-and-effect relationship with the US election. However, before considering  causation, we need to rule out alternative explanations:

\vspace{1pt}\noindent \emph{New users.} If new users are suddenly introduced in the sample, heart rate volume could increase. That does not apply to our case since, for the whole duration of the year, we study the very same set of users whose heart rate is monitored almost continuously throughout the year (90\%+ of the days).

\vspace{1pt}\noindent \emph{Software/hardware update.} Device and  software updates might impact measurements. During the year of observation, our devices' software that measured heart rates did not change, and all measurements showed high consistency. 

\vspace{1pt}\noindent \emph{Physical Activity.} Heart rates could increase as a result of increased physical activity. However, there was no substantial change in daily number of steps at the time of the US election (a person did, on average, 6794 steps a day in October, 6750 in November, and 6660 in December).

\begin{figure}[t!]
\centering
\includegraphics[width=.45\columnwidth]{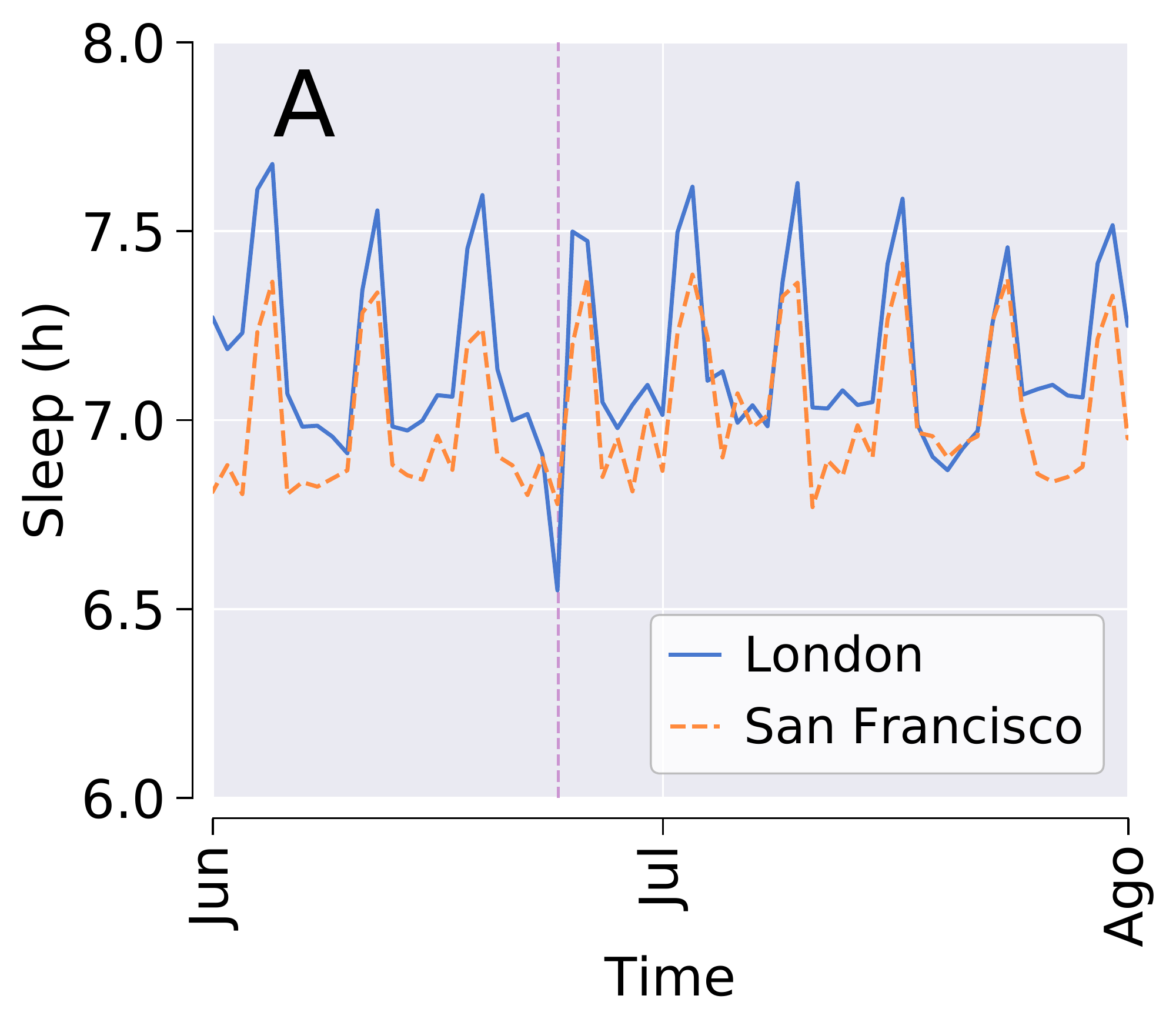}
\includegraphics[width=.45\columnwidth]{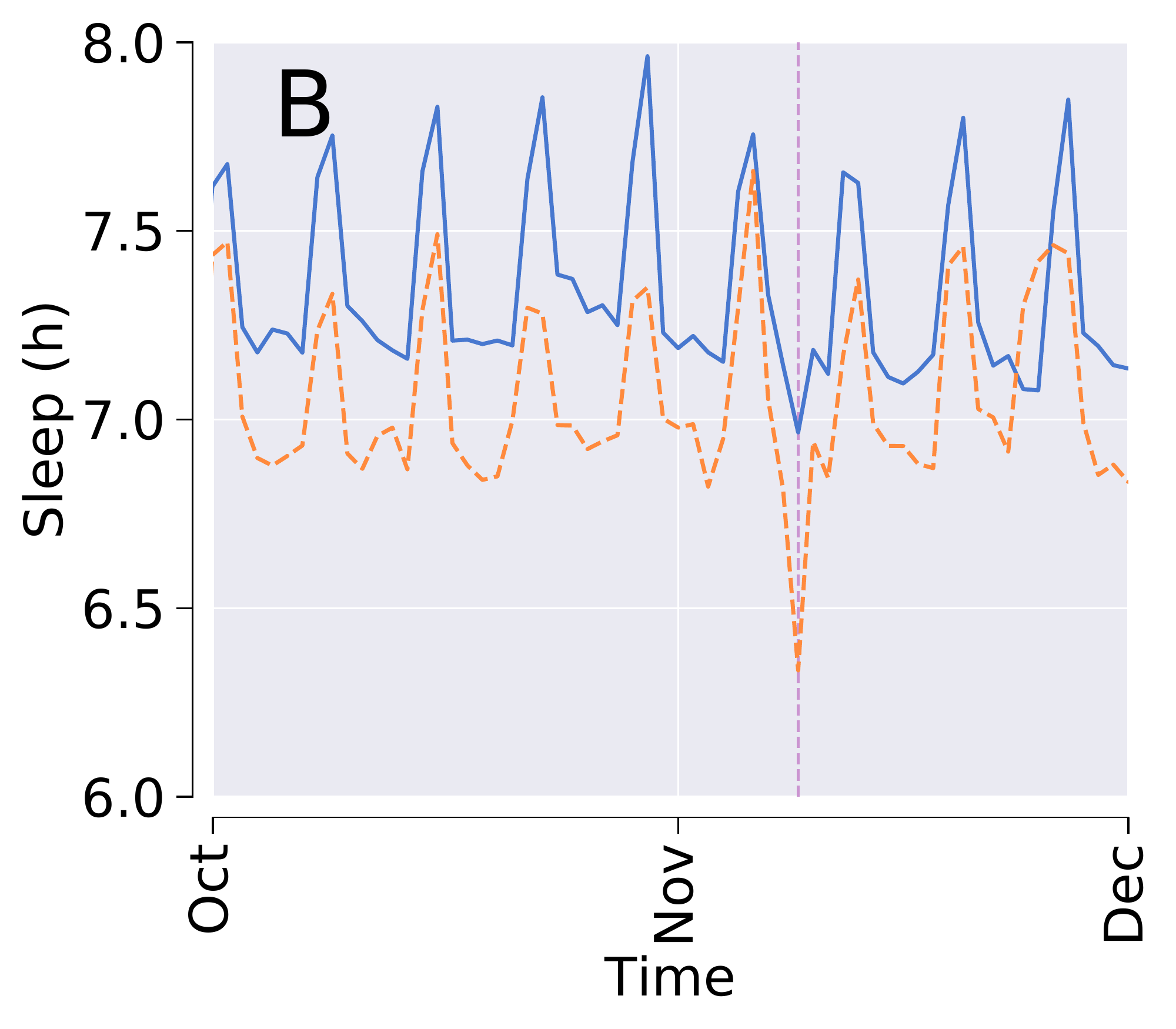}\\
\includegraphics[width=.45\columnwidth]{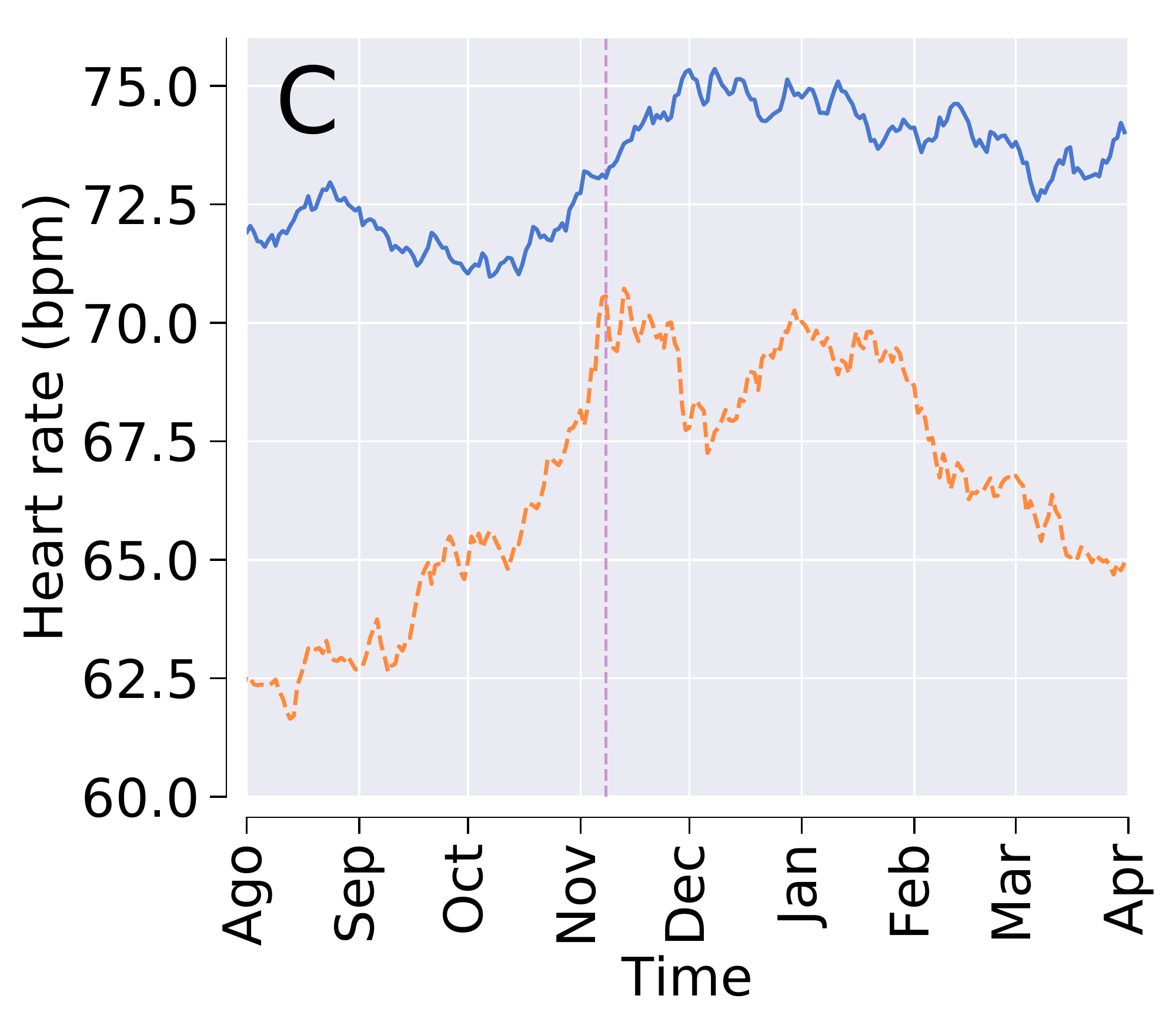}
\includegraphics[width=.45\columnwidth]{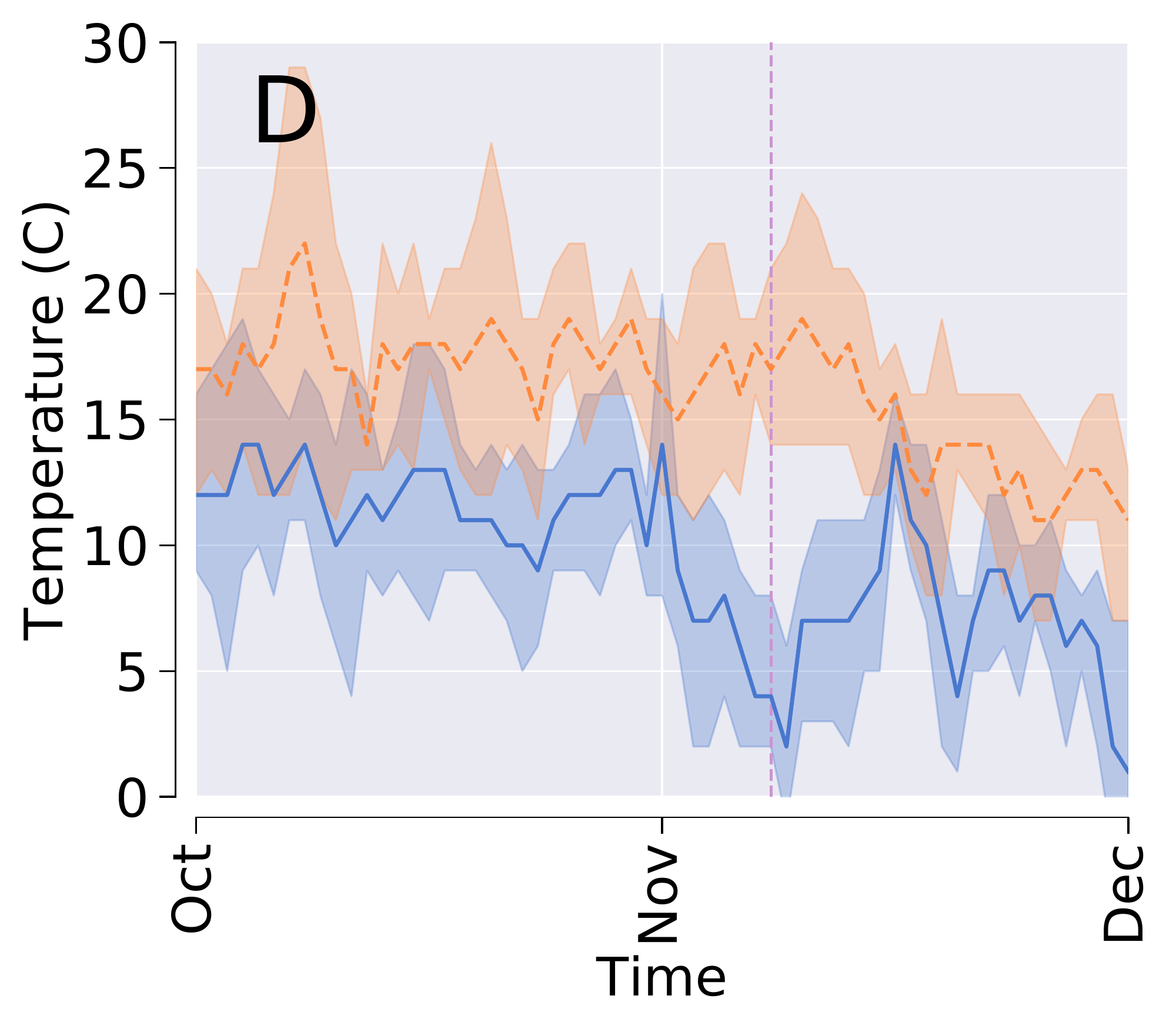}\\
\caption{(A) Number of hours slept in the weeks around Brexit. (B) Number of hours slept for the US presidential election. (C) Average heart rate around the US presidential election. (D) Temperature changes for the US presidential election (the average temperature curve is enclosed within the \emph{min} and \emph{max} curves).}
\label{fig:confounding}
\end{figure}

\vspace{1pt}\noindent \emph{Temperature.} In cold weather, to keep the body warm, the heart beats faster. The temperature in the months concerning the US election was stable (Figure~\ref{fig:confounding}D), ruling out temperature as co-founding factor.

\vspace{1pt}\noindent \emph{Seasonality.} People's rhythms are seasonal~\cite{golder11}. However the observed heart rate increases are steady and are not seasonal. If they were, given the comparable weather conditions (Figure~\ref{fig:volume}),  the heart rate levels in April 2016 would be the comparable to those in April 2017 -- but they are not.

Upon observational data, it is hard to argue what caused heart rate increases. However, the strongest association appears to be with the US election, and that is because of three  main reasons:

\vspace{1pt}\noindent \emph{(i) Alternative Explanations.} We have just ruled out the most plausible explanations other than the US election. 

\vspace{1pt}\noindent \emph{(ii) External Validity}. Increases in heart rates have been found to be associated with emotional regulation and stress~\cite{vrijkotte2000effects,hjortskov2004effect,thayer2012meta}. It should come as no surprise that the US election \emph{caused} (self-reported) stress in a considerable part of the electorate.  Based on a representative sample on 1,000+ US residents, a survey commissioned by the American Psychological Association found that more than half of the interviewees experienced the political climate around the presidential campaign as a significant source of stress~\cite{apa2016stress}. 

\vspace{1pt}\noindent \emph{(iii) Dose-response relationships}. Dose-response patterns on observational data are necessary (but not sufficient) for considering causation. In our case, we indeed observe that events are strongly linked to biorhythm responses. To see how, contrast Londoners with San Franciscans:  San Franciscans experienced rapid heart rate increases  the last two months of the US political campaign (Figure~\ref{fig:confounding}C),  experienced a peak exactly on the election day, and slept the least  during the US election night (Figure~\ref{fig:confounding}B); by contrast,  Londoners slept less the night after Brexit (Figure~\ref{fig:confounding}A), and started to experience heart rate increases on the US election day (Figure~\ref{fig:confounding}C), suggesting that their response was shifted compared to the US counterpart, as one expects. Therefore, dose-response relationships are observed for both the US election and Brexit. 

\section{Synchronicity} \label{sec:synchronicity}

So far we have captured the volume of steps, hours slept, and heart rates. To go beyond volume, we now focus on temporal patterns.  The timing of behavior has always been a strong expression of the style of individuals and entire populations~\cite{lynch1972time}. Nowadays that timing can be reliably captured by smart devices. Our sleep data, for example, includes the time at which users go to bed every day. This can be interpreted as an ordered sequence of timestamps, which is also called \textit{spike train}. For the purpose of this study, we are interested in measuring the degree of synchronization between two users, that is, between two spike trains $s_1 =\{t_1^{(1)}, ..., t_n^{(1)}\}$ and $s_2 =\{t_1^{(2)}, ..., t_m^{(2)}\}$, within an interval $[0,T]$ of, say, one year. The SPIKE-distance function $D_S$ provides a parameter-free way of doing that~\cite{mulansky2015guide}. It is defined as the integration of an instantaneous spike function $S(t)$ over time: $D_S(0,T) = \frac{1}{T} \int_{0}^{T} S(t) dt \in [0,1]$.

The spike function at time $t$ is defined as:
\begin{equation}
S(t) = \frac{ |\Delta t_P(t)| \cdot \langle x_F^{(n)}(t) \rangle_n +  |\Delta t_F(t)| \cdot \langle x_P^{(n)}(t) \rangle_n}{ \langle ISI \rangle_n }
\label{eq:spike_function}
\end{equation}
where $\Delta t_P(t)$ is the difference between the two spikes $t_P^{(1)}(t)$ and $t_P^{(2)}(t)$ that immediately precede time $t$ in the two trains; $\Delta t_F(t)$ is the difference between the spikes following $t$; $x_P^{(n)}(t)$ is the distance between $t$ and the previous spike in the $n^{th}$ train; $ISI^{(n)}$ is the mean inter-spike interval in the $n^{th}$ train; and $\langle \bullet \rangle_n$ denotes the average over the two trains. When $D_S=0$, the two trains have no distance between them, meaning that their spikes are perfectly synchronized; when $D_S=1$, the two trains are completely out-of-phase. The formulation of $D_S$ for the bivariate distance (for 2 users) can be extended to a multivariate case (for 2+ users) by averaging the distances of all the pairs of spike trains in the set. We compute that quantity and denote it with $\overline{D_S}$.

\begin{figure}[t!]
\centering
\includegraphics[width=.99\columnwidth]{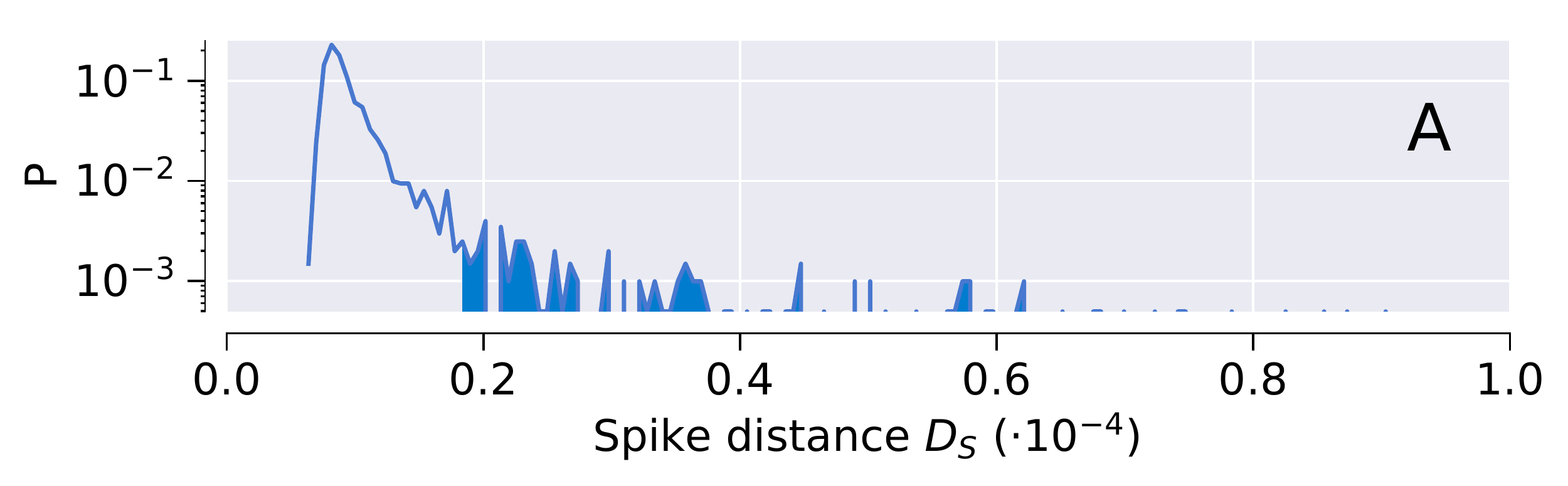}\\
\includegraphics[width=.49\columnwidth]{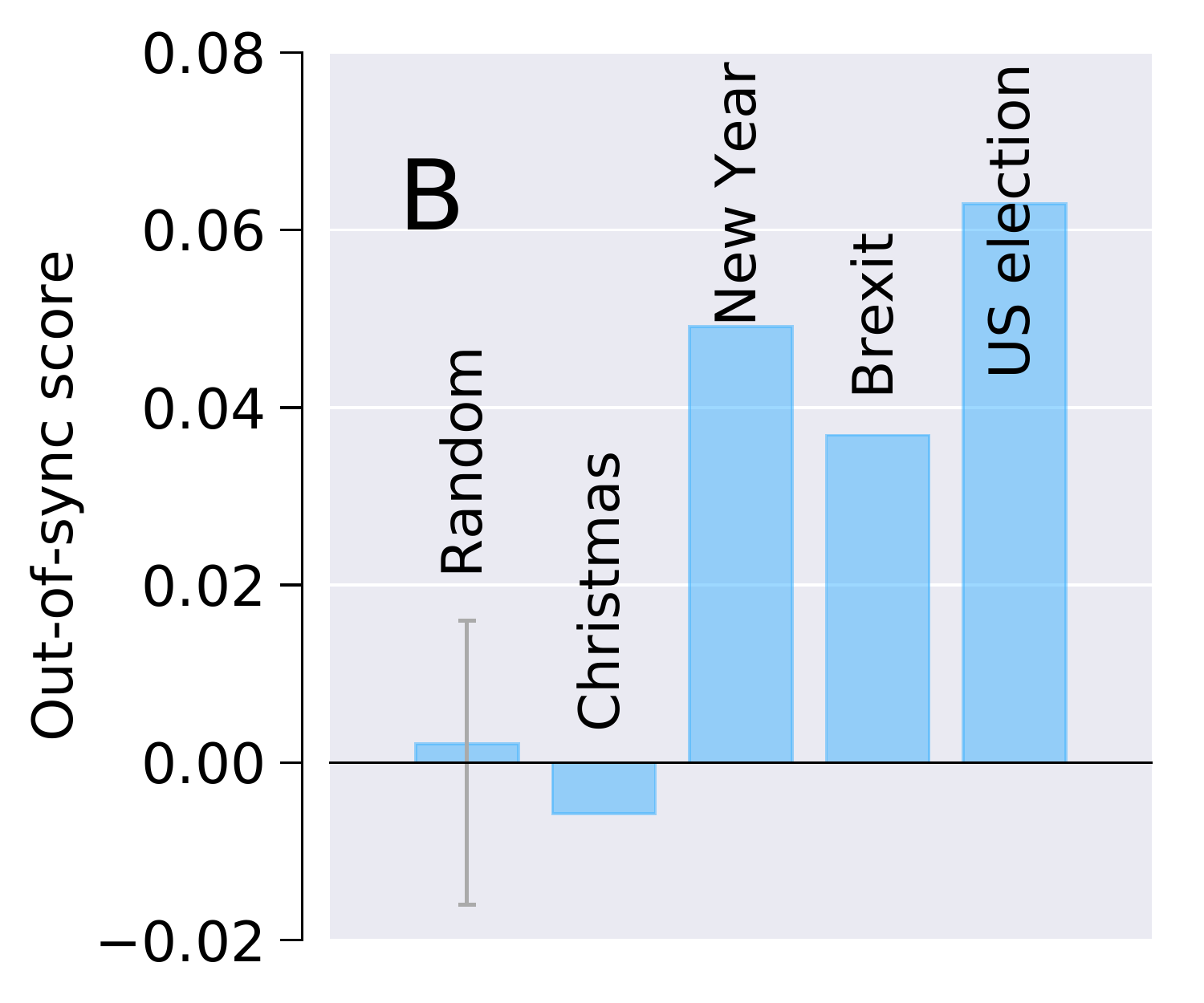}
\includegraphics[width=.49\columnwidth]{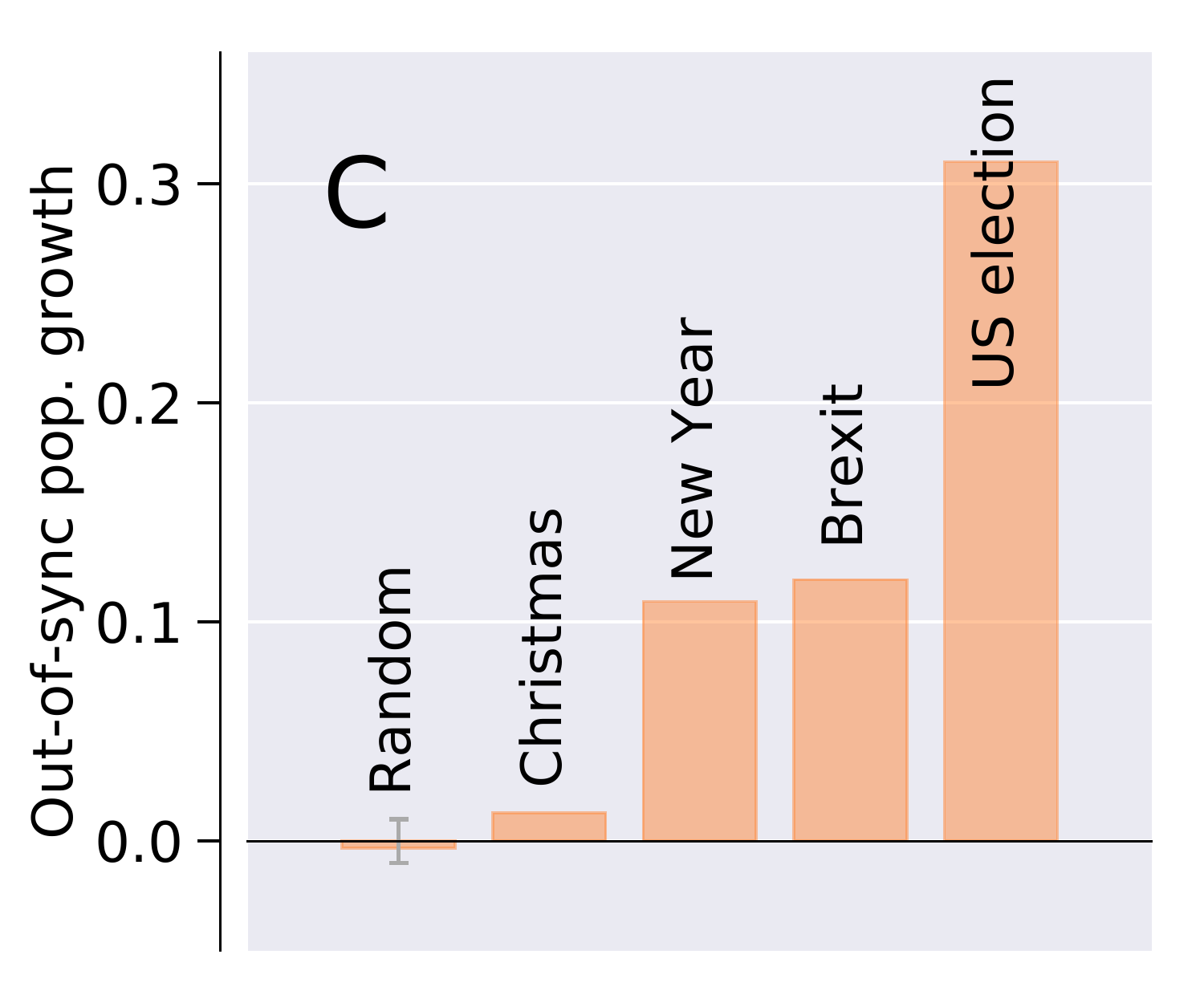}
\caption{(A) Frequency distribution of the spike distance $D_S$ over the whole population during the full year. The tail (the set of points that are 2 standard deviations away from the median) is marked with a solid blue area under the curve. (B) Out-Of-Sync score ($OOS$) for the four events plus the random model computed with a $95\%$ confidence interval. (C) Out-Of-Sync population growth ($OOS^{\uparrow}$) for the four events  plus the random model computed with a $95\%$ confidence interval. }
\label{fig:synchronicity}
\end{figure}

Each user's sleep patterns for the entire year have been converted into a spike train. This consists of the sequences of times at which the user went to sleep. The level of de-synchronization in the population is then computed as the average spike distance score $\overline{D_S}$ over all user pairs. Even if theoretically bounded in $[0,1]$, the $D_S$ variable takes values from $0$ to $10^{-4}$ in our data (Figure~\ref{fig:synchronicity}A). That is because $\overline{D_S}$ is quite low for events influenced by exogenous events (e.g., it is rare to find a considerable number of people who sleep in the middle of the afternoon). To quantify the extent to which  synchronization changes after each of our four events, we compare the average spike distance $\overline{D_S}$ over all user pairs  in the week before the event, and in the week after the event.  More formally, we define the \textit{Out-Of-Sync score} ($OOS$) of an event $e$ occurring at time $t$ as $OOS_{e@t}= \overline{D_S}(t, t+\alpha)-\overline{D_S}(t-\alpha,t)$,

where $t$ is the time of event $e$, and $\alpha$ is a buffer time window, which, in our case, we set to be of  one week. If the event's score is above zero, then this means that, after the event, the population became, on average, less synchronized. If it is below zero, then the population became more synchronized. To make sure that an event's score is not due to chance, we contrast it to a null/random model, that is, we contrast it to what the score would be if computed at random days. More specifically, we compute $OOS$ with $\alpha$ of one week at 100 random days along the whole year, obtaining 100 scores. We then average those scores out to obtain the random model's score, which is supposed to be zero. By definition, the accuracy of the $D_S$ measure (and, consequently, that of $OOS$) suffers in the presence of missing data points, which is the case for our data, since our devices are not perfectly reliable. As such,  to get robust measurements, we filter out all users whose spike trains are not complete in the weeks before each event, and in the weeks after it. This step turns out to exclude at most a few hundred individuals for each event.

After this filtering, we compute the out-of-synch scores $OOS$. We find that, at random days, the scores are close to $0$, as one expects. By contrast, the scores are subject to changes during three of our four events. More specifically, they do not change in a statistically significant way during Christmas, but they do considerably change during New Year's Eve, Brexit, and the US election, suggesting that several users became out-of-sync after these three events (Figure~\ref{fig:synchronicity}B). To quantify the fraction of the population who slipped considerably out-of-sync after each event, we consider the frequency distribution of out-of-sync scores (Figure~\ref{fig:synchronicity}A): its right tail represents those user pairs who are heavily out-of-sync with each other. Using a standard practice in outlier identification~\cite{leys2013detecting}, we consider  all the points that are at least 2 standard deviations (2$\sigma$) higher than the median ($\tilde{\mu}$) as outliers: $\textrm{outliers}(D_S(t,t+\alpha)) = \int_{\tilde{\mu}+2\sigma}^{1}f_{D_S(t,t+\alpha)}(x) dx$,

where $\alpha$ is the considered time window  (i.e., one week), and $f$ is the probability density function computed for the variable $D_S$ in the time period $[t,t+\alpha]$. To then measure the impact of an event $e$, we compute the value for the previous expression of $\textrm{outliers}()$ after $e$ minus its value before $e$, and normalize the result:
\begin{equation}
OOS^{\uparrow}_{e@t} = \frac{\textrm{outliers}(D_S(t,t+\alpha))-\textrm{outliers}(D_S(t,t-\alpha))}{\textrm{outliers}(D_S(t,t-\alpha))}
\label{eq:pop_out_of_synch_increase}
\end{equation}
The resulting value is the \textit{Out-Of-Synch population growth} ($OOS^{\uparrow}$): it is the relative increase in the portion of user pairs that are heavily out-of-sync. From Figure~\ref{fig:synchronicity}C, one sees that the value increased by 10\% after New Year's Eve and Brexit, and by as much as 30\% after the US election. The random baseline shows no increase. 


\section{Rhythms} \label{sec:rhythms}

As a final metric, we consider circadian rhythm. This is a roughly 24 hour cycle in the physiological processes of living beings, including humans. Although circadian rhythms are endogenous (``built-in''), they are adjusted to the local environment by external cues such as light and temperature. Disruptions in a person's circadian rhythm  for sleep and heart rates have been found to have negative health consequences~\cite{ritcher60} and lead to pathological conditions~\cite{saurbh11, takahashi08}.  To see how to track  circadian rhythms on our data of sleep patterns and heart rates, consider that any activity signal over time can be interpreted as a \textit{time series}, an ordered sequence of activity measurements. To extract the period of an activity time series, one can use the Discrete Fourier Transform. This decomposes the temporal signal into a number of discrete frequencies which, if recombined, compose the original signal. The \textit{Power Spectral Density (PSD)} is the distribution of relative power of those frequencies; we extract it using the Welch method~\cite{welch1967use}. To make the results more interpretable, we transform the frequencies of the PSD into periods ($period(PSD)$), which denote the amplitudes of the wave originated by those frequencies, expressed in number of days (e.g., a period of 7 days denotes a weekly pattern). More formally, given a user $u$'s time series in a period $[t_1, t_2]$, we define its characteristic \textit{rhythm} as the period with the maximum PSD value in $[t_1, t_2]$: $rhythm_{u}(t_1, t_2) = \argmax(period(PSD(t_1,t_2)))$.

Our goal is to go beyond individual users and quantify the \emph{rhythm shift} associated with an event in the entire population. To this end, for any given event $e$ that took place at time $t$, we first compute the rhythm shift an individual user $u$ experienced before and after the event, within a temporal window $\alpha$: $\textrm{rhythm shift}_{u,e@t} = rhythm_{u}(t, t+\alpha) - rhythm_{u}(t-\alpha,t)$.

We then aggregate the rhythm shift values across all users by computing their frequency distribution $f_{\textrm{rhythm shift}_{e@t}}$. To make sure our shift values are not due to chance, we resort to a null/random model. We compute such model's score value by computing ``rhythm shift'' scores for 100 random days: we first compute \emph{individual} shift scores around those days ($\textrm{rhythm shift}_{u,rand@t}$), and then compute the distribution over all users and days ($f_{\textrm{rhythm shift}_{rand@t}}$).

Finally, to estimate the entire population's rhythm disruption associated with $e$, we compare the observed distribution for $e$ with the distribution for random days: $\textrm{rhythm disruption}_{e@t}= D_{KL}(f_{\textrm{rhythm shift}_{e@t}}, f_{\textrm{rhythm shift}_{rand@t}})$. This is the KL divergence between the two frequency distributions~\cite{KL}. The higher it is, the higher the rhythm shift that is associated with the event compared to random expectation (zero for no shift).

\begin{figure}[t!]
\centering
\includegraphics[width=.49\columnwidth]{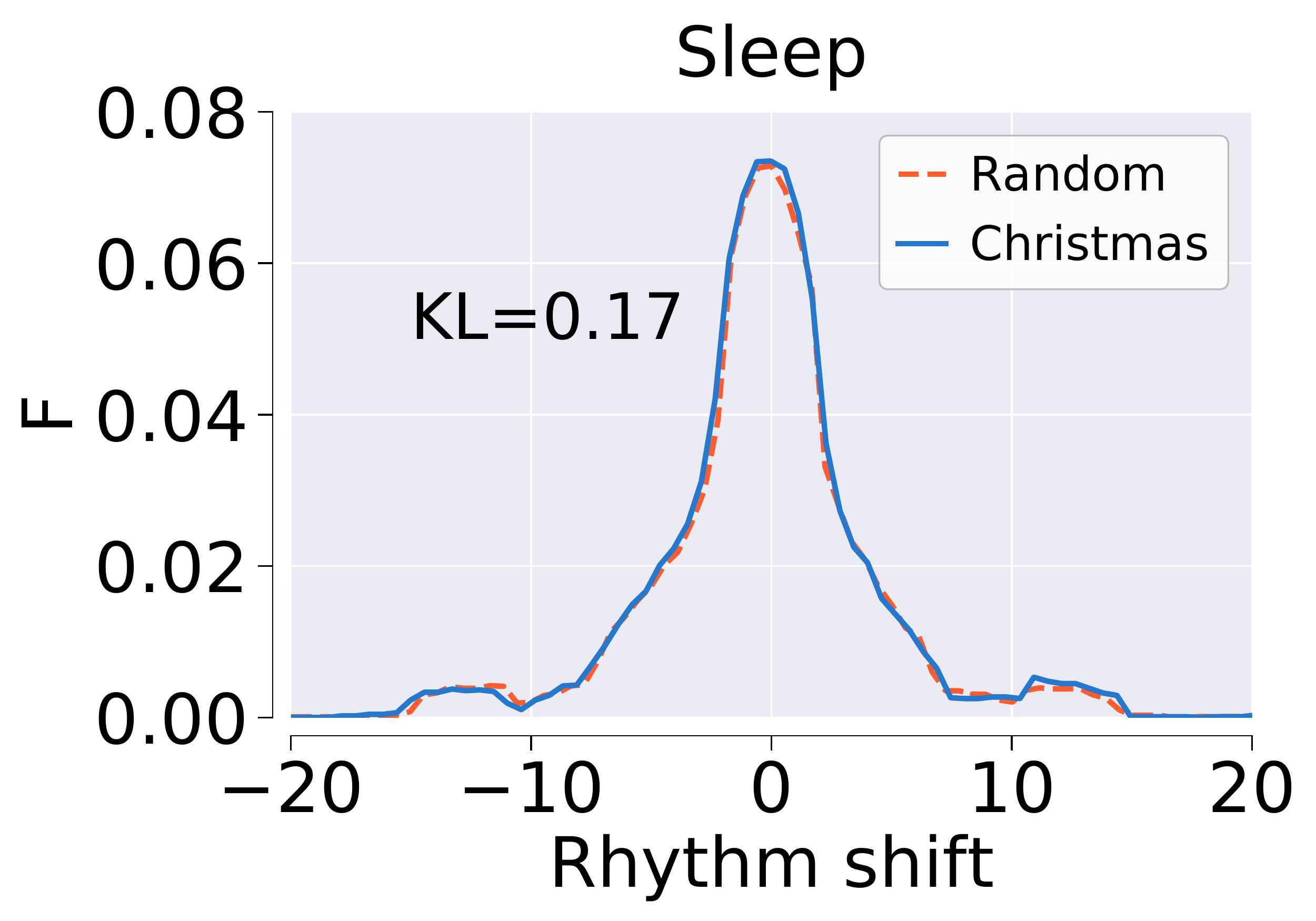}
\includegraphics[width=.49\columnwidth]{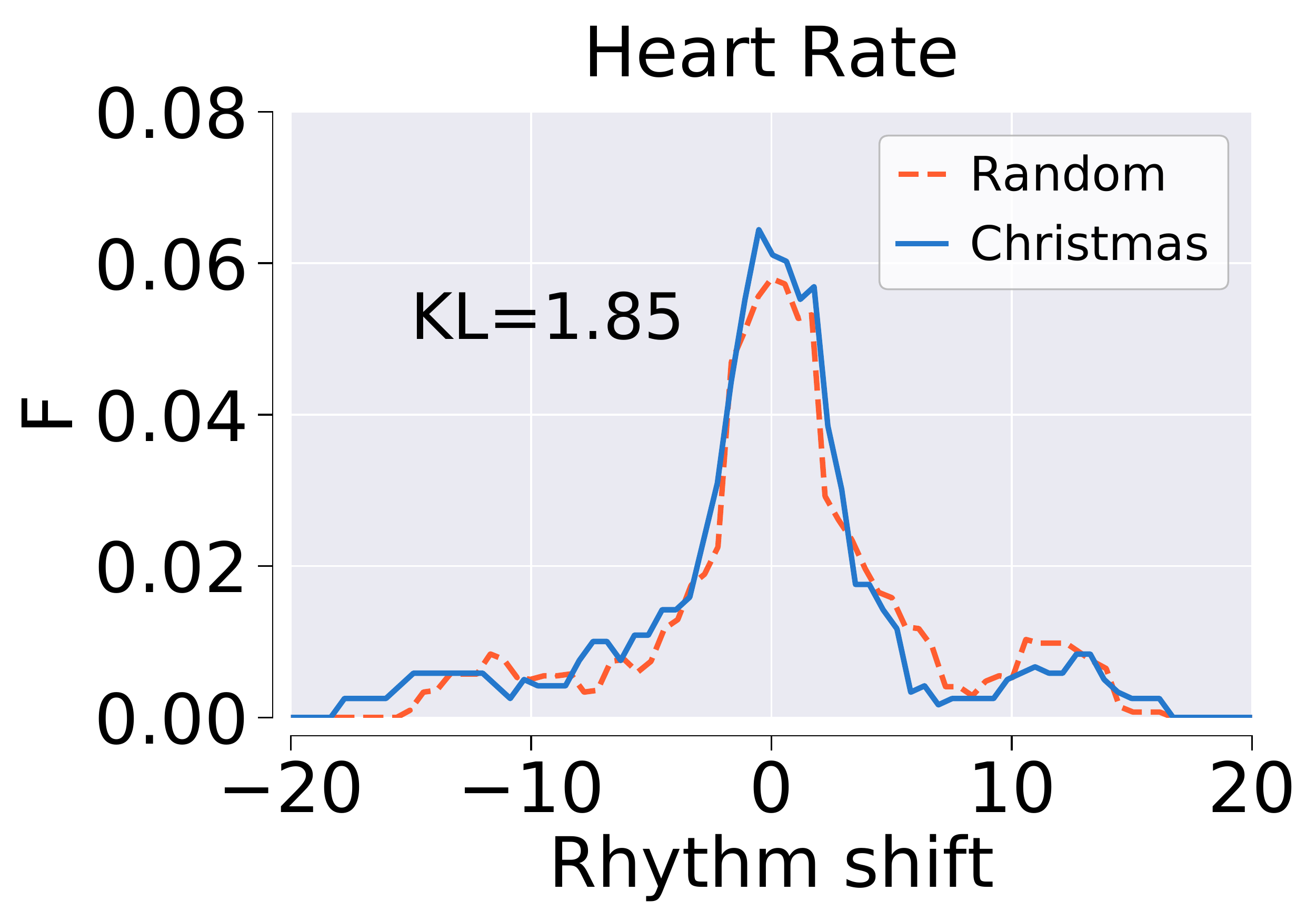}\\
\includegraphics[width=.49\columnwidth]{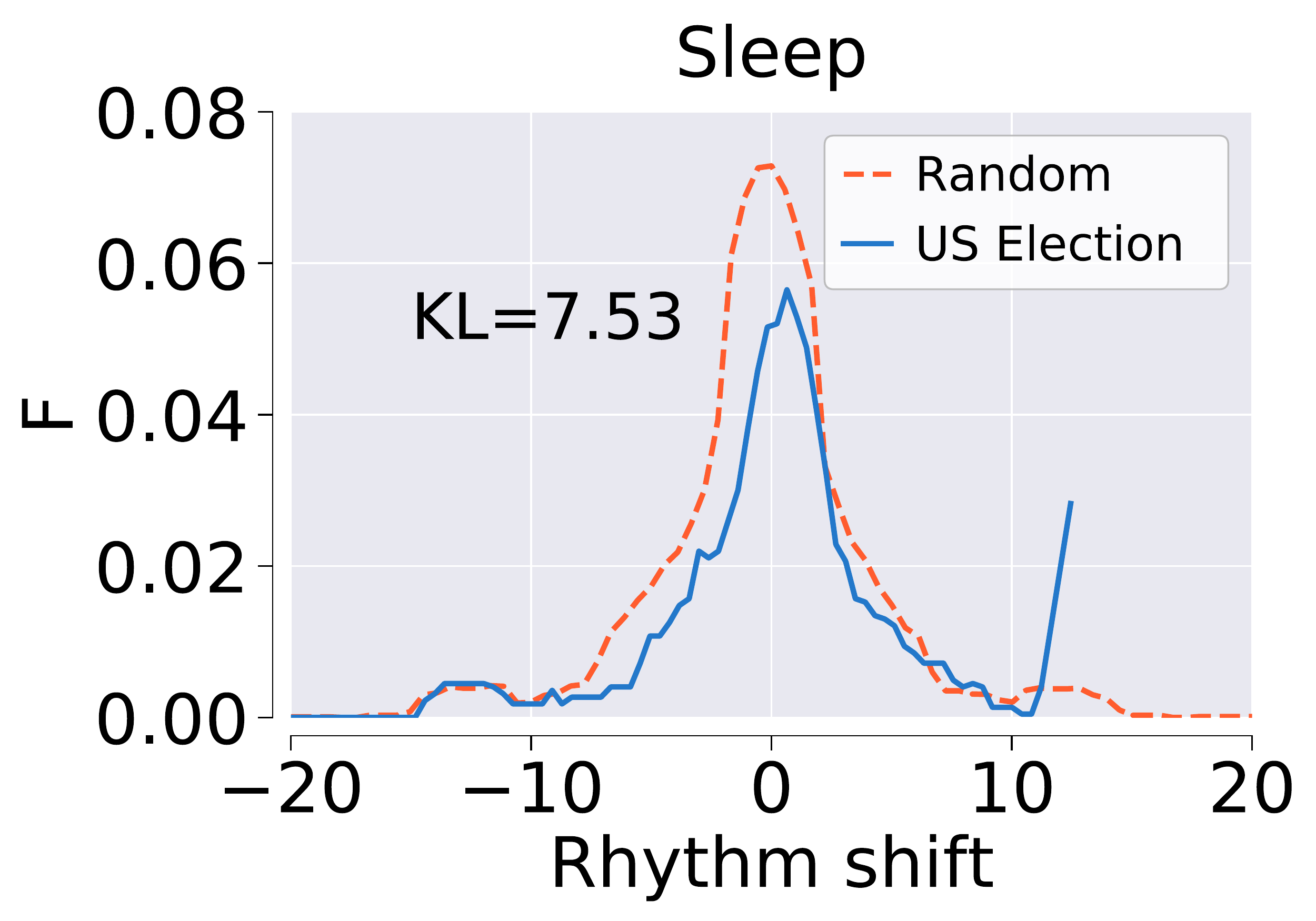}
\includegraphics[width=.49\columnwidth]{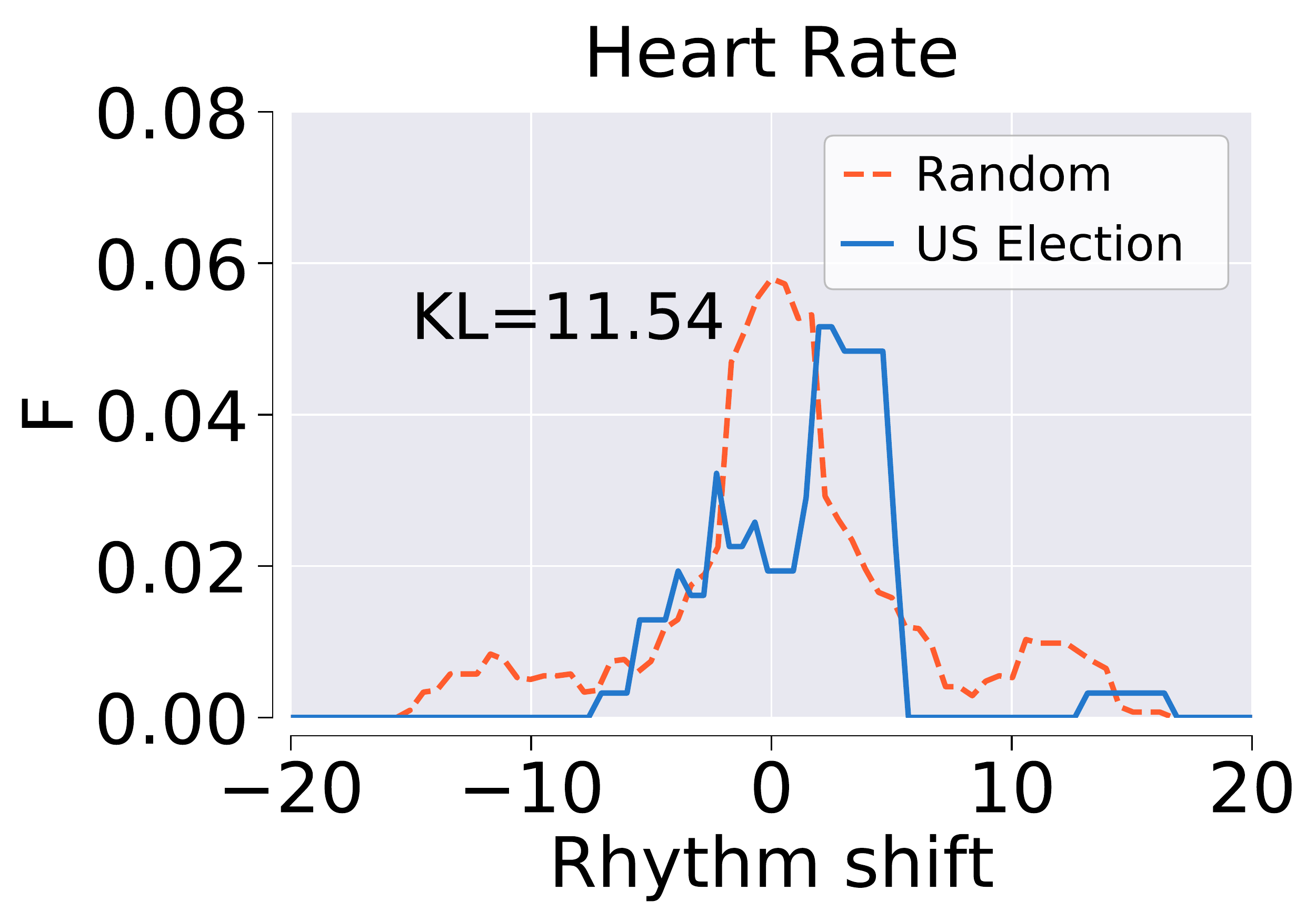}
\caption{Frequency distributions of the rhythm shifts for sleep and heart rates after Christmas and the US presidential election. The KL-divergence between the observed distribution and the random model's is our \textit{rhythm disruption}.}
\label{fig:periodicity}
\vspace{-4pt}
\end{figure}

Figure~\ref{fig:periodicity} shows the distribution of \textrm{rhythm shift} for sleep and heart rates around Christmas and the US election. Each distribution is shown together with the corresponding null/random model's distribution, and the difference between the observed distribution and the random one (called `rhythm disruption')  is also reported and denoted with ``KL''.  After Christmas and New Year's Eve, the shifts for both sleep and heart rates are limited. Instead, after the US election and Brexit, the shift for sleep is considerable, and that for heart rates is disruptive\footnote{Due to page restrictions, we show the results for Christmas and the US election here and invite the reader to visit \url{http://goodcitylife.org/health} for more.}. 


\section{Conclusion}

\newcolumntype{C}[1]{>{\centering\arraybackslash}p{#1}}
\newcolumntype{L}[1]{>{\raggedright\arraybackslash}p{#1}}
\begin{table}[t]
\centering
\begin{tabular}{L{1.4cm}|C{0.4cm}C{0.4cm}C{0.6cm}|C{1.0cm}|C{0.6cm}|C{0.4cm}C{0.4cm}C{0.5cm}}
\textbf{Event} & \multicolumn{3}{c|}{\textbf{Volume}} & \textbf{OOS}  & \textbf{OOS$^{\uparrow}$} &  \multicolumn{3}{C{2cm}}{\textbf{Rhythm Disruption}} \\
\hline
\hline
                     & Steps & Sleep & Heart & Sleep & Sleep & Steps & Sleep & Heart \\
\cline{1-9}
\textit{Brexit}      & 7564 & 6.6 & 69.6 & 37$\cdot 10^{-3}$ & 12\% & 0.06 & 0.37 & 9.61 \\
\textit{US}          & 7042 & 6.7 & 70.7 & 62$\cdot 10^{-3}$ & 32\% & 0.43 & 7.53 & 11.5 \\
\textit{Christmas}   & 4531 & 7.7 & 71.5 & -6$\cdot 10^{-3}$ & 10\% & 0.13 & 0.17 & 1.85 \\
\textit{New Year's}  & 5370 & 6.9 & 71.1 & 49$\cdot 10^{-3}$ &  1\% & 0.12 & 0.17 & 1.64 \\
\textit{Random}      & 7129 & 7.2 & 71.4 & 49$\cdot 10^{-3}$ &  0\% &  0.0 &  0.0 & 0.0  \\
\end{tabular}
\caption{Average values for each of our metrics for our four events.}
\label{table:values4metrics}
\end{table}

\begin{figure}[t!]
\vspace{-10pt}
\centering
\includegraphics[width=.49\columnwidth]{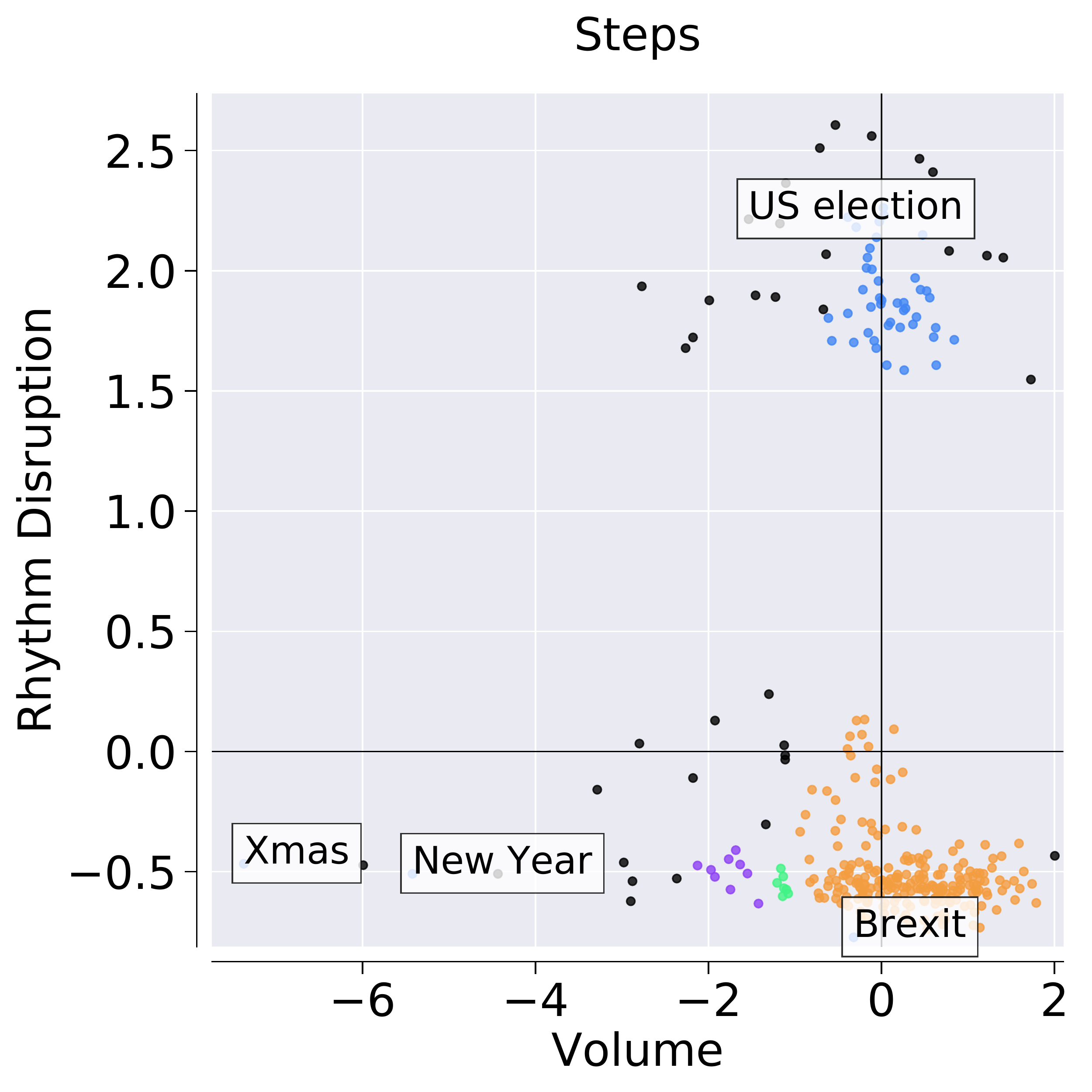} 
\includegraphics[width=.49\columnwidth]{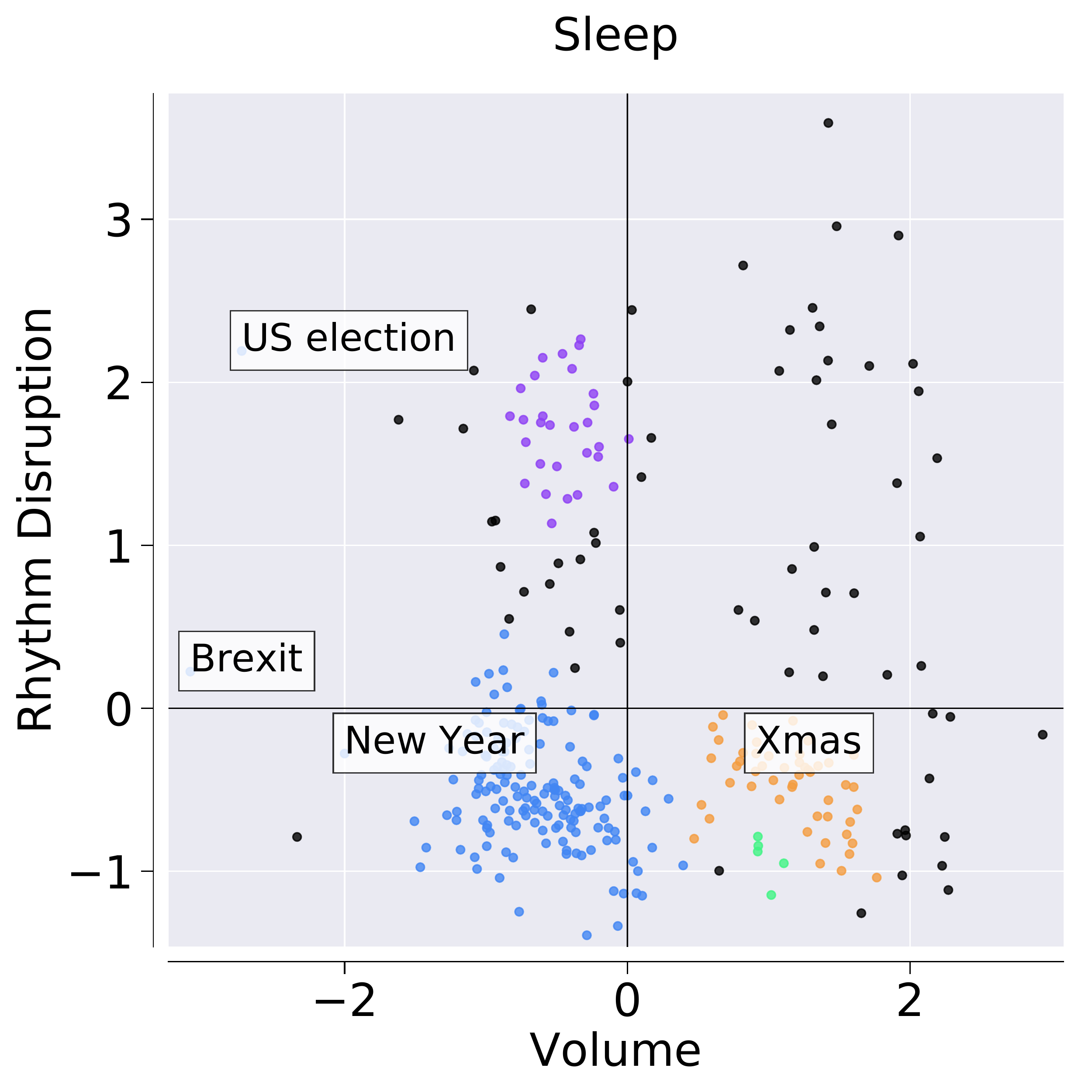} 
\caption{Clusters of the days of the year under study based on the volume and rhythm disruption for steps and sleep. Each dot corresponds to a day, and the colors encode the different (DBSCAN) clusters (black dots are outliers and, as such, do not belong to any cluster). The days in which our four main events happened are marked with labels. Values on both axes are standardized.}
\vspace{-10pt}
\label{fig:quadrants}
\end{figure}

Based on all the results, one might hypothesize that each of our metrics could offer a  way of profiling large-scale events. In reality, no individual metric considered in separation would be sufficient. For example, from Table~\ref{table:values4metrics}, one can see that volume alone is not a reliable marker for distinguishing the four events under consideration: Brexit is indistinguishable from New Year's Eve, for instance. By contrast, considering our metrics in combination is sufficient for distinguishing the four events. Indeed, by plotting the daily average number of steps in a ``rhythm disruption by volume'' plot (Figure~\ref{fig:quadrants}A), the four events are separable (i.e., they form distinctive clusters), suggesting that rhythm disruption and volume are, in our case, reliable  markers for event classification.  The same applies to daily average number of hours slept (Figure~\ref{fig:quadrants}B). This is further supported by a DBSCAN clustering of those points, which returns a silhouette value (clustering quality value~\cite{rousseeuw1987silhouettes}) of $\sim$0.35.

Taken all together, our results are very promising, yet three main limitations hold. First, our users are not representative of the  general population. Second, our metrics suffer from data sparsity and, to be generalizable, they need to be furthered researched. Finally, our results do not speak to causation.  Still, despite those limitations and based on the dose-response nature of the relationships between  events and biorhythm measurements, we can conclude that, with our metrics at hand, one is able to capture ``how we experience time'' in unobtrusive ways. Synchronization and rhythms seem to be present in all living beings. They generally serve to keep the inner organisms working and keep the body coordinated with the external world. A failure of synchronization puts the body out-of-sync and under stress. Nowadays smart health trackers are able to capture that experience, and are able to do so at an unprecedented scale. 

\bibliographystyle{ACM-Reference-Format}

\end{document}